\documentclass[a4paper,twoside]{article}

\usepackage{epsfig}
\usepackage{subfigure}
\usepackage{calc}
\usepackage{amssymb}
\usepackage{amstext}
\usepackage{amsmath}
\usepackage{amsthm}
\usepackage{multicol}
\usepackage{pslatex}
\usepackage{apalike}

\graphicspath{{./figs/}}

\usepackage{array}
\newcolumntype{P}[1]{>{\centering\arraybackslash}p{#1}}

\usepackage{url}
\usepackage{enumitem}
\usepackage{dblfloatfix}
\usepackage{SCITEPRESS}     

\makeindex

\begin{document}
\title{Industrial Survey on Robustness Testing In Cyber Physical Systems}

\author{Christophe Ponsard, Abiola Paterne Chokki, and Jean-François Daune
\affiliation{CETIC Research Centre, Charleroi, Belgium}
\email{\{christophe.ponsard, abiola.chokki, jean-francois.daune\}@cetic.be}}


\abstract{Cyber-Physical Systems (CPS) play a critical role in modern industrial domains, including manufacturing, energy, transportation, and healthcare, where they enable automation, optimization, and real-time decision-making. Ensuring the robustness of these systems is paramount, as failures can have significant economic, operational, and safety consequences. This paper present findings from an industrial survey conducted in Wallonia, covering a wide range of sectors, to assess the current state of practice in CPS robustness. It investigates robustness from how it is understood and applied in relationship with requirements engineering, system design, test execution, failure modes, and available tools. It identifies key challenges and gaps between industry practices and state-of-the-art methodologies. Additionally, it compares our findings with similar industrial surveys from the literature.
}

\onecolumn \maketitle \normalsize \vfill

\section{\uppercase{Introduction}}

In many application fields, such as rail systems, automotive, intelligent buildings and Industry 4.0, the development of specific systems is raising more and more challenges, not least due to their complexity, but also to the interactions between hardware, software and network components. These systems, known as cyber-physical systems (CPS), need to operate robustly in open, dynamic environments that are potentially exposed to malicious actors. 

Robust operation is a major issue, impacting not only development costs and timescales, but also the company's reputation, and is often dealt with on an ad hoc basis, with highly variable practices. It affects both large companies integrating complex systems (e.g. automotive, railways) and SMEs developing more specific products that are integrated into their customers' solutions or environment (e.g. environmental sensors, instruments for a space platform).

The context of our work is a Walloon project aiming to help companies, and more specifically SMEs, ensure a higher degree of robustness for their increasingly software-based products and services, particularly complex cyber-physical systems or products integrated into systems of systems \cite{carapace}. The goal is to offer them a toolbox to support robust design and operation activities with an agile approach aligned with DevSecOps practices and with a good level of automation, including through the use of robustness test generation using model-based/mutation testing \cite{Hierons21} and fuzzing \cite{Manes19}. It also aims to rely on fault injection practices such as Chaos Engineering popular in Cloud native environment \cite{Rosenthal20}. One of the first steps in this project was to study the state of practices and needs in this field, by means of a survey which is the scope of this paper. Our work was primarily designed to guide our project but can also be useful to share to a broader audience and to cross-check if others have reported similar observations from the industry.

In order to reach this goal, this paper is structured as follows. First, Section 2 details the survey process including the questionnaire design which is aligned with a previous survey to enable comparability of results. Section 3 characterises the surveyed companies. Section 4 provides a detailed analysis of our results. Section 5 compares them with some similar surveys and studies. Finally Section 6 concludes and identifies our future work.

\section{\uppercase{Survey Design \& Process}}

In order to ensure the collection process described in the previous point, the questionnaire was designed with the following requirements in mind:
\begin{itemize}[noitemsep]
\item gather information on needs and practices for CPS robustness
\item collect key company information (size, field of activity, type of software engineering activity) to characterize needs and practices
\item not take too long to fill (about 30 min)
\item be sufficiently framed by giving suggested answers, which allows for better comparability and also the possibility of giving non-anticipated choices
\item be compatible with surveys conducted elsewhere on similar themes, thus enabling comparisons to be made on a broader basis
\end{itemize}

With regard to the last point, we were able to identify some approaches in the literature for carrying out a state-of-practice study. The most elaborate and documented is a Swedish study carried out in 2016 \cite{Shah16}. The questionnaire was found to be very interesting to reuse in terms of its overall structure, which adopts an approach covering the logic of handling robustness throughout the life cycle of a CPS, from requirement statement, design, testing and monitoring in production. It also covers the tooling aspect. The overall structure is detailed hereafter with comparable questions marked in italic.
\begin{itemize}[noitemsep]
\item definitions
\begin{itemize}[noitemsep]
\item agreement on \textit{robustness} and CPS definitions
\item \textit{company or domain specificities}
\end{itemize}
\item robustness requirements
\begin{itemize}[noitemsep]
\item \textit{source}
\item \textit{examples}
\end{itemize}
\item robustness design practices
\item performing robustness testing
\begin{itemize}[noitemsep]
\item when/\textit{how}/\textit{who}
\item \textit{environment}
\item \textit{indicators}
\item specificities
\end{itemize}
\item robustness failure
\begin{itemize}[noitemsep]
\item \textit{classification} (CRASH imposed)
\item \textit{root cause analysis}
\item actions for tracking, elimination, detection
\end{itemize}
\item robustness tooling
\begin{itemize}[noitemsep]
\item \textit{used/known tools}
\item \textit{most useful tool functions} 
\item \textit{missing tool functions}
\end{itemize}
\end{itemize}

It should be noted that these questions are fairly open-ended, as the questionnaire was primarily  used in semi-structured interviews, following a similar protocol that the Swedish study. In our case, each company interview lasted for about 1h30. The questionnaire was available a few days before under the form of a online (google) form which could be prefilled. \textbf{The full questionnaire is available online in English version} \cite{form}.

Most of the contacted companies contacted are SMEs which had confirmed their interest in joining the project advisory board. Some extra companies were also identified later and were also proposed to join the advisory board.

\section{\uppercase{Characterisation of the companies}}

\begin{figure}[h]
\centering
\includegraphics[width=\columnwidth]{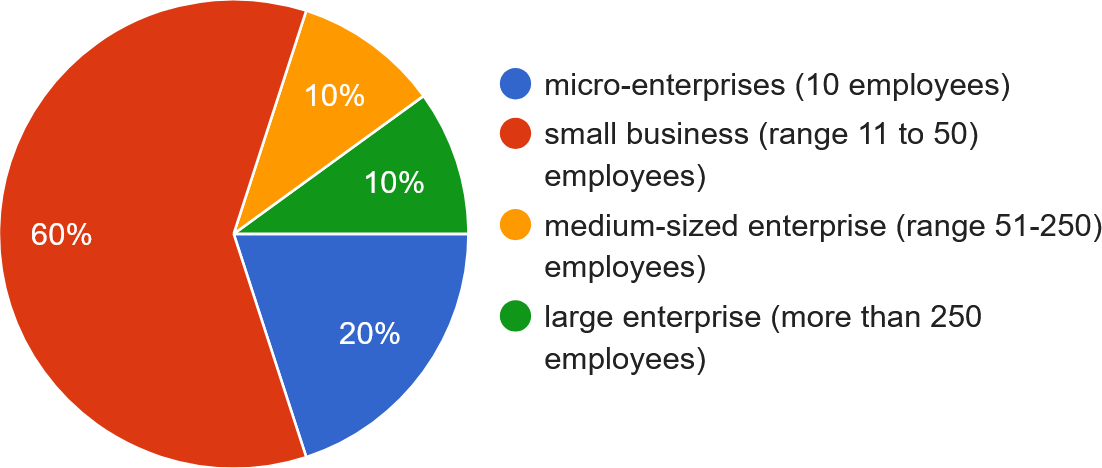}
\caption{Company size distribution}
\label{fig:sizes}
\end{figure}

A total of 10 companies were surveyed using the protocol described in Section 2 which is of the same order as the 13 surveyed in \cite{Shah16}. The surveyed companies were mostly SMEs, with only one large company. Most have between 11 and 50 employees (including one start-up). One very small and one medium-sized company were also contacted. This distribution depicted in Figure \ref{fig:sizes} is fairly representative of the Walloon landscape \cite{stats}.

\begin{figure}[h]
\centering
\includegraphics[width=0.9\columnwidth]{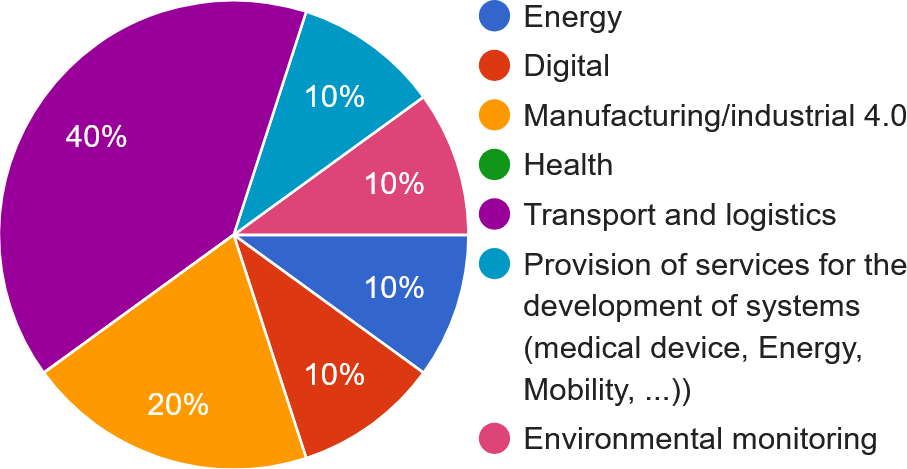}
\caption{Surveyed sectors}
\label{fig:sectors}
\end{figure}

As shown in Figure \ref{fig:sectors}, the survey has a good coverage of the industrial sectors  with some more representatives from transport/logistics and manufacturing/industry 4.0. A missing sector is healthcare, which is however addressed by one of the digital companies as a customer.

\begin{figure}[b!]
\centering
\includegraphics[width=\columnwidth]{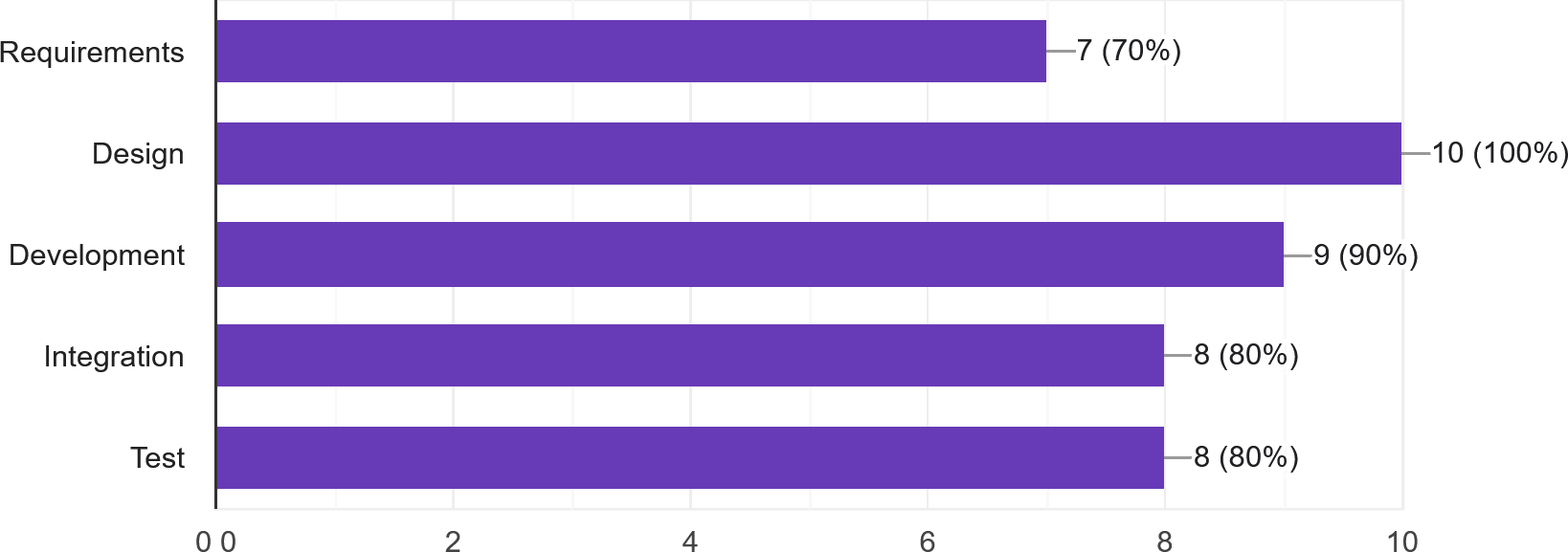}
\caption{Activities carried out by companies}
\label{fig:lifecycle}
\end{figure}

In terms of software lifecycle activities, these are all covered, from specification to testing, as illustrated in the Figure \ref{fig:lifecycle}. Most companies cover the entire lifecycle and develop end-to-end CPSs, even if in certain scenarios some are suppliers of partial solutions integrated by others. Two companies are more focused: one specializes solely in requirements engineering, and covers only the specification and high-level design aspects. Another specializes in specific sensors that are integrated into solutions developed by others.

\section{\uppercase{Detailed Analysis}}

We provide here a detailed analysis based on the structure presented in Section 2.

\subsection{Definition and Issues}

With regard to the definition of robustness, all companies agreed with the IEEE definition of "the degree to which a system or component can function correctly in the presence of invalid inputs or stressful environmental conditions" \cite{ISO24765_2017}, sometimes with certain nuances and clarifications such as the importance of ensuring continuity of operation despite a deteriorating environment, monitoring the level of availability of system resources, and analyzing the possibility that disruptions are of malicious origin. This last aspect of cybersecurity was unanimously and spontaneously reported as a major concern by all companies.

The companies also agreed on the concept of CPS and the fact they were building or integrating some form or CPS. Two key characteristics of a CPS were frequently mentioned: (1) reliability: resistance to erroneous requests, reliability of sensors over time, problem detection (non-nominal system) and (2) cybersecurity concerns: resistance to malicious actions, especially those targeting system availability or integrity.


\subsection{Robustness Requirements and Design Practices}

As far as robustness requirements are concerned, they mainly come from customers and internal practices. Standards are relatively rarely cited. Customers are more likely to express SLA (Service Level Agreement) requirements in terms of performance, reactivity, availability and load, which ultimately concern robustness capabilities. Within the company, a certain culture may be present through the more systematic collection of robustness requirements alongside other non-functional requirements by means of a document template. Companies subject to certification constraints in terms of operational safety have a much higher level of maturity from the outset throughout the development cycle, and in particular for requirements, by explicitly specifying normal and degraded modes of operation.

\begin{figure}[h]
\centering
\includegraphics[width=\columnwidth]{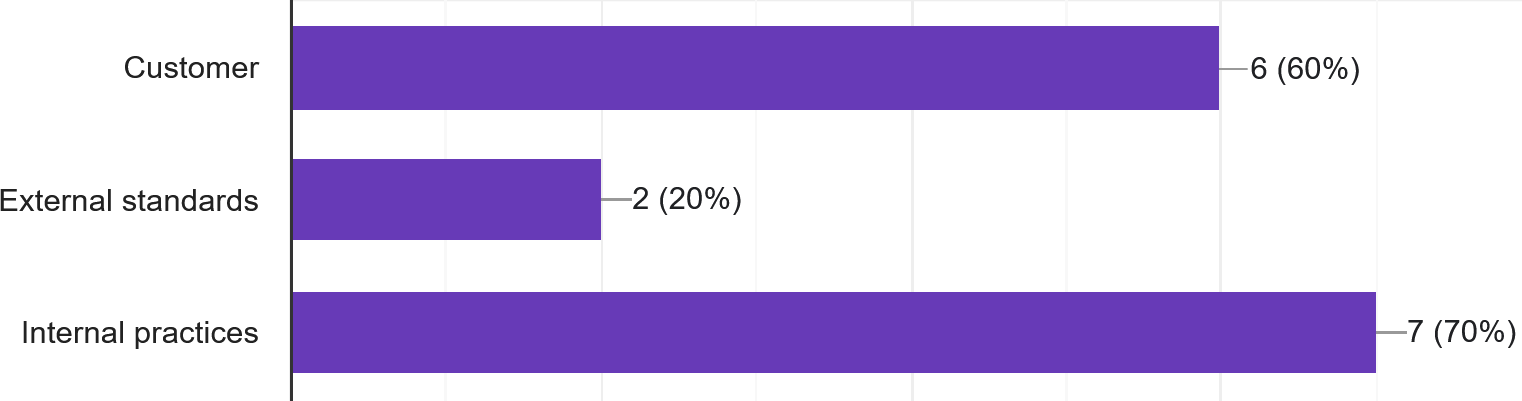}
\caption{Origin of robustness requirements}
\label{fig:reqs}
\end{figure}

\begin{figure*}[b!]
\centering
\includegraphics[width=0.9\textwidth]{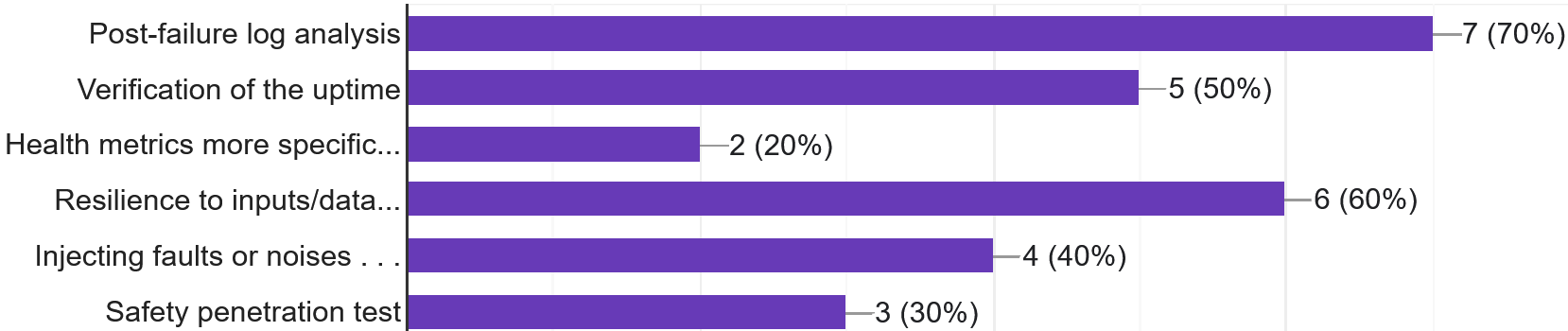}
\caption{Robustness measurements}
\label{fig:indicators}
\end{figure*}

Some typical robustness requirements, grouped by category, are as follows:
\begin{itemize}[noitemsep]
\item existence of a degraded mode (cited by 2 companies with a high level of maturity):
    \begin{itemize}[noitemsep]
    \item degraded mode operation on architectures with several redundant rail processing chains, without impact on reaction times even when switching between modes
    \item the X system will support a degraded mode in the event of a malfunction (an offline mode is cited, with a reduction to a subset of the system's most critical functions; the need for a resynchronization process is also identified).
    \item presence of a manual procedure (in a limited way, for an information system that are ready for it, typically with a resynchronisation procedure when back to full operation)
    \end{itemize}
\item expression of certain SLAs to be ensured, particularly for:
    \begin{itemize}[noitemsep]
    \item availability: in terms of SIL standards or more specifically via downtime (e.g. linked to restart procedures).
    \item reactivity/response times, sometimes hyper-reactive (nanoseconds on some medical , on the order of a hundred ms in railway applications) 
    \item need for measurement accuracy and maintenance over time (cited by two companies, particularly the one active in sensors)
    \item continuous running time: able to run for a week without failure
    \end{itemize}
\item design constraints such as limit CPU consumption by X\% 
\end{itemize}


With regard to robust design and development practices, the main measures implemented during the design phase are as follows:
\begin{itemize}[noitemsep]
\item use of a priority mechanism for critical functions
\item implementation of redundancy, detection and recovery mechanisms
\item micro-services architecture or with low coupling
\item installation of an overlay to compensate for drift (e.g. sensors)
\item backup and restoration plan
\end{itemize}
And concerning the measures implemented during development :
\begin{itemize}[noitemsep]
\item systematic validation of entries
\item exception handling
\item assessing the robustness of 3rd-party components
\item peer review
\end{itemize}

\subsection{Robustness Testing}

Robustness tests are generally carried out at the end of development, after sufficient functional validation has been achieved at component or system level on nominal cases, and before the deployment phase. One company puts forward the implementation of tests pushing the system to its limits to ensure that it is capable of performing diagnostics even if the tests are not yet fully playable. In the context of agile development, one practice mentioned is to include robustness tests every 2 or 3 sprints and at the end of the project. Scenarios linked to updates were also mentioned.

The tests implemented cover several categories: long-term tests in the lab and also in the field, load tests aimed at reaching the system's limits by calling on as many functions as possible at the same time, hardware failure simulation tests, failover tests, interruption and recovery tests.

As for the role to which robustness testing (and more broadly, testing) is entrusted, in the majority of cases it's a shared responsibility within the development team (generally following the cross-developer/independent tester pattern) or, for structures with more elaborate teams, a multi-skilled tester. It should be noted that there is no specific profile for a robustness tester: the role is both too specialized and requires functional knowledge of the system to be tested.

\begin{figure}[h]
\centering
\includegraphics[width=\columnwidth]{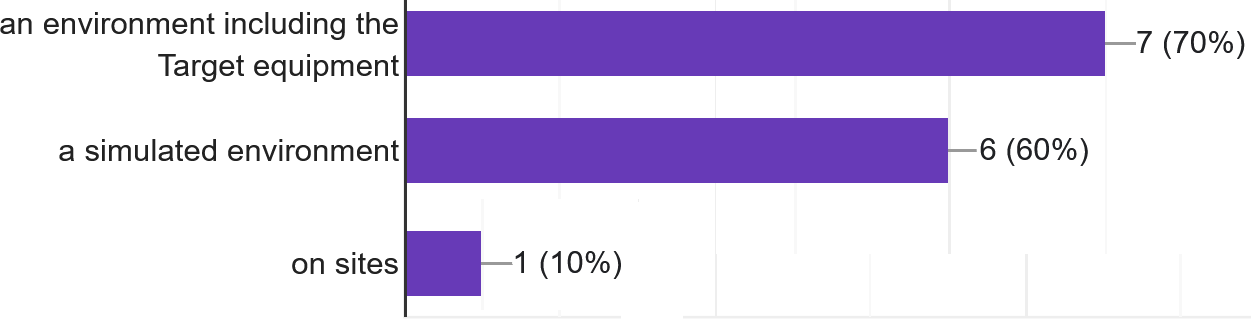}
\caption{Testing environment}
\label{fig:env}
\end{figure}

In the vast majority of cases, the test environment is mixed, with a simulated component and a target hardware component as depicted in Figure \ref{fig:env}. The general aim is to be able to carry out as many tests as possible in the controlled environment of the lab, realistically simulating the real environment, with a good level of automation and controlled costs. On-site tests are carried out to validate the real environment, but are much more costly and therefore minimized.

An important element of strategy is therefore the development of the test environment with the following aspects:
\begin{itemize}[noitemsep]
\item be sufficiently realistic while limiting costs (the upper limit being the cost of testing on site)
\item the development of a test environment is a fairly complex task that requires a good command of it. One company mentioned that it preferred to have complete control over this type of environment, rather than use one supplied even in part by a third party.
\item the scope of the environment is determined by the context and constraints (cost, timing). For example, you can limit yourself to a purely "batch" environment if the user interface aspect is not relevant. Otherwise, you may need to consider the simulation or integration of an interface component in the test environment.
\item we can also capitalize more if there is a product (line) dimension versus an approach focused on a customer's specific (one-shot) project
\end{itemize}

In the operational monitoring phase, Figure \ref{fig:indicators} shows the most frequently examined elements are logs of key system components, resilience to erroneous input/output and uptime. Other more specific elements are noise resistance on hardware component signals, penetration tests (specific to security) and certain status metrics (CPU occupancy, current state of a PLC, etc.) System health factors are quite rarely defined.

\subsection{Robustness Failure}

Operational failures can be classified according to the CRASH terminology (Catastrophic, Restart, Abort, Silent/unobservable, Hindering), originally developed in the context of Operating Systems \cite{CRASH}. All types were reported by the companies, with only one catastrophic case. One difficulty reported several times concerned the problem of diagnosing failures that are not very reproducible (Hindering), or in the case of Silent.

The following causes have been identified as possible causes of failure:
\begin{itemize}[noitemsep]
\item defects in functional specifications (incompleteness, lack of precision, particularly in interaction with other systems in the environment)
\item non-functional specifications that do not specify robustness-related elements
\item poor implementation of specifications (bugs)
\item non-robust architecture problems
\item poor exception handling
\item cybersecurity issue causing data or code alteration
\item problems linked to faulty interactions with equipment or the environment
\end{itemize}

\noindent The main cause of a failure is identified via:
\begin{itemize}[noitemsep]
\item analysis of fault and log reports
\item investigation and analysis of dependencies
\item the (attempted) reproduction of the failure
\item methodical problem isolation via step-by-step debugging, hypothesis testing, possibly systematized by division ("binary search")
\item examination of hardware/software against a reference "probe
\end{itemize}

\noindent In order to detect similar problems, the following measures are suggested:
\begin{itemize}[noitemsep]
\item impact analysis (possibly to identify changes to be made to the entire downstream branch of the "V" development model)
\item enrichment of regression tests
\item verification of the test coverage of the code concerned
\item targeted reinforcement of code reviews
\item general improvement of V\&V and QA processes
\end{itemize}

\noindent In terms of the correction process and avoiding their repetition in the future
\begin{itemize}[noitemsep]
\item eliminate the root cause and validate the correction
\item workarounds are to be avoided, but in a context where everything is not under control, architectures with monitoring and recovery mechanisms can be used (e.g.watchdog + restart).
\item in general: hardening of system architecture
\item document the problem and its correction, share knowledge
\item improve test planning
\end{itemize}

\noindent Main means mentioned for early fault detection before it results in a failure are:
\begin{itemize}[noitemsep]
\item in the design phase: design review
\item in the coding phase: code review
\item in the test phase: continuous testing, test coverage
\item at runtime: monitoring system, automated and/or human alert/exception management, more comprehensive laboratory testing
\item at global level: risk management and continuous quality improvement processes
\end{itemize}

\subsection{Robustness Tooling}

As far as the tools used for robustness management are concerned, the situation is quite varied, but generally speaking, we can make the following observations:
\begin{itemize}[noitemsep]
\item there is no specific tool for design or preparation.
\item generic tools are used, based on the battery of tools available for other types of testing.
\item in the most elaborate cases, a test environment can be used to carry out robustness testing actions, such as fault injection, simulation of component failure or unexpected environmental stress.
\item verification is largely based on logs, but can be carried out ad hoc using in-house tools or by exploiting certain dedicated tools (e.g. wireshark for network trace exploration).
\end{itemize}

\noindent Tooling needs can largely depend on the context, but here are some suggested elements:
\begin{itemize}[noitemsep]
\item use templates to specify robustness
\item combinatorial management: generate/simulate "all" scenarios (not only functional, but also combinations of illicit and erroneous data)
\item automation: continuous testing (including long-term testing)
\item monitoring: the problem of unified to merge various traces/logs timestamping 
\item time management: estimation of time required, prioritization
\item load tests
\item broader suggestion: investigate the possible contributions of AI
\end{itemize}

\section{\uppercase{Comparative discussion with related surveys}}

As stated before, our survey was inspired by a related industrial survey carried out in the scope of testing industrial embedded systems \cite{Shah16}. Looking at the respective results of analysis on each topic, we can identify a number of similarities and some deviations:
\begin{itemize}[noitemsep]
\item about robustness requirements, the companies in both surveys have similar comments about the definition, stressing that automatic and quick recovery mechanisms is a form of robustness or the presence of a safe/degraded mode (especially in critical system context). 
\item the case of deliberate attack (cybersecurity) seems to have evolved from many companies to a vast majority in our case. 
\item availability is a key requirement in both surveys with also secondary requirements about latency and responsiveness.
\item the origin of requirements is also mixed but more internal when client are less mature and more from the client for more mature or regulatory in specific domains, typically involving safety (mentioned in transportation in our case).
\item the classification of the organisation is less elaborated in our survey as no organisation had a dedicated staff for robustness testing and in only bigger companies there is actually dedicated staff for testing. In most SMEs, testing is organised inside the development team with some level of formalisation of the procedure. 
\item crash failure was only mentioned once in our context but the question was formulated about what was observed in operation rather than during test sessions. In our context, the main concern was diagnosability related to silent or hindering failures.
\item about the tooling, the situation is quite similar with an in-house environment built in an ad hoc way out of built-in and open-source tools. The expectations about ideal tools are also converging about more automated and end-to-end tools covering the execution of test scenarios, the monitoring and identification of the test outcome.
\end{itemize}

Another interesting study focused only on robustness testing in the telecom domain (Ericsson), also in Sweden \cite{Eldh12}. The focus is a large telecom system. It also involved structured interviews but also inspection and observation. The context can only be compared with the big company involved in our own survey but highlight a number of challenges of similar nature such as the manual nature of the creation and selection of test cases, the difficulty to create a representative traffic load, the analysis and qualification of external failure reports from customers, and also determining when a system version is stable enough to go into robustness testing given the release deadline.

A complementary view is given in the CPS-focused survey by \cite{Gunes14}. It covers a variety of similar sectors as our study like manufacturing, transportation, healthcare, critical infrastructures. Robustness is one of the key challenges. Common to our survey, it identified possible disturbances possibly arising from sensor noises, actuator inaccuracies, faulty communication channels, potential hardware errors or software bugs. Moreover, it highlights the lack of modelling of integrated system dynamics, such as real ambient conditions in which CPSs operate, evolving operational environments, and unforeseen events. These factors, which may be unavoidable at runtime, emphasise the need for robust CPS design.

Finally, a more recent study focuses on the use of logs in embedded software engineering which is also relevant for CPS \cite{Yang23}. It relies on interview studies to highlight that text-based editors and self-made scripts are still commonly used when it comes to tooling in log analysis practice. The quality of logs is also a challenge to be addressed by log instrumentation and management.

\section{\uppercase{Conclusion and Perspectives}}

This paper presented a survey work about robustness testing of Cyber Physical Systems carried out in Belgium to guide the elaboration of adequate support to help SMEs developing such systems. To sum up the main identified needs in robustness testing of CPS, most companies stressed the complexity of understanding the CPS to be tested, often combined with incomplete documentation of certain system components. This difficulty can be exacerbated by architecture evolutions, sometimes introduced late in the project and constituting a risk. Another factor is the cumbersome nature of robustness testing compared with other types of testing, such as unit testing. Especially, it involves the necessity to set up an elaborated test environment, which must be close to the real system and require the implementation of simulated parts. It is also difficult to diagnose sometimes hard to reproduce problems and identify the root cause. In addition, taking care of possible malicious causes requires enlarge robustness with cybersecurity considerations.

Our survey was also discussed in the light of other industrial studies which have confirmed the majority of observations and challenges that we identified  across the whole robustness testing lifecycle from robustness requirements identification to testing and beyond to runtime monitoring in operation. 

Those elements will now feed the actual work that is carried out with a number of the interviewed SMEs. In order to be agile and support automation, our research line is to base our work on Chaos Engineering practices which provide a systematic way of identifying assumptions, defining test experiments, carry them out using specific tools to identify how robust the system is, then improve it as required and replay the process in an automated way. It also provides a variety of tools for performing fault injection and highly effective distributed monitoring. Those practices are however currently more widely used for Cloud native environment and need some adaptation for CPS which is the next step of our research.

\noindent \paragraph{Acknowledgements.}

This survey was partly funded by project CARAPACE (grant 2310088). Warm thanks to the SMEs who took part in our survey.

\bibliographystyle{apalike}
\bibliography{icsoft25survey}

\end{document}